\documentclass[letterpaper,11pt, preprint]{elsarticle}
\usepackage{easybmat}
\usepackage{tabularx}
\usepackage{etex}
\usepackage{tikz}
\usetikzlibrary{matrix,arrows,decorations.pathmorphing}

\usepackage{bbm}
\usepackage[english]{babel}
\usepackage[T1]{fontenc}
\usepackage[ansinew]{inputenc}

\usepackage{lmodern}	
\usepackage{amsmath}	
\usepackage{amsthm}
\usepackage{amsfonts}
\usepackage{mathtools}
\usepackage{amssymb}
\usepackage[amssymb]{SIunits}
\usepackage{nomencl}
\usepackage[version=3]{mhchem}
\usepackage[super]{nth}
\usepackage{enumerate}
\usepackage{graphicx}

\RequirePackage{ifthen}
\renewcommand{\nomgroup}[1]{%
 \ifthenelse{\equal{#1}{V}}{\item[\textbf{Variables}]}{%
 \ifthenelse{\equal{#1}{P}}{\item[\textbf{Problem Parameters}]}{%
 \ifthenelse{\equal{#1}{C}}{\item[\textbf{Physical Constants}]}{%
 \ifthenelse{\equal{#1}{D}}{\item[\textbf{Derived Constants}]}{%
 \ifthenelse{\equal{#1}{S}}{\item[\textbf{Scaling Units}]}{%
 \ifthenelse{\equal{#1}{N}}{\item[\textbf{Numbers}]}{%
 \ifthenelse{\equal{#1}{F}}{\item[\textbf{Functions}]}{%
 \ifthenelse{\equal{#1}{U}}{\item[\textbf{Super- and Sub-Scripts}]}{%
}%
}%
}%
}%
}%
}%
}%
}%
}
{\left\{\begin{tabular}{>{$}r<{{}$}@{}>{${}}l<{$}r>{$}r<{{}$}@{}>{${}}l<{$}}}%
{\end{tabular}\right.}

\makenomenclature

\usetikzlibrary{arrows,shapes,backgrounds}
\tikzstyle{every picture}+=[remember picture, baseline=0]
\tikzstyle{matnode} =[inner sep=0pt,outer sep=3pt,anchor=base,remember picture]
\tikzstyle{matline} = [overlay,draw,loosely dotted,thick]

\makeatletter
\def\rdots{\mathinner{\mkern1mu\raise\p@
\vbox{\kern7\p@\hbox{.}}\mkern2mu
\raise4\p@\hbox{.}\mkern2mu\raise7\p@\hbox{.}\mkern1mu}}
\makeatother

\newcommand{\prn}[1]{{\left(#1\right)}}
\newcommand{\brk}[1]{{\left[#1\right]}}
\newcommand{\brc}[1]{{\left\{#1\right\}}}
\newcommand{\fprn}[1]{{\!\prn{#1}}}
\newcommand{\fbrk}[1]{{\!\brk{#1}}}
\newcommand{\abs}[1]{{\left|#1\right|}}
\newcommand{\norm}[1]{{\lVert#1\rVert}}

\newcommand{\oneover}[1]{\frac{1}{#1}}
\newcommand{\FD}{{\mathrm{FD}}}
\newcommand{\fFD}{{f^\FD}}
\newcommand{\hatfFD}{{\hat f^\FD}}
\newcommand{\fzero}{{f^{(0)}}}
\newcommand{\hatfzero}{{\hat f^{(0)}}}
\newcommand{\nui}{{\nu_\text{i}}}
\newcommand{\nue}{{\nu_\text{e}}}
\newcommand{\goto}{{\rightarrow}}

\begin{document}
\begin{frontmatter}

\title{Charge transport in a super\-lattice: a numerical study using moment methods}
\author{Yossi Farjoun \corref{cor1}}
\ead{yfarjoun@ing.uc3m.es}
\cortext[cor1]{Corresponding author.}
\author{Luis L. Bonilla}
\ead{bonilla@ing.uc3m.es}
 
\address{Gregorio Mill\'an Institute for Fluid Dynamics, Nanoscience and Industrial Mathematics, Universidad Carlos III de Madrid,
Avenida de la Universidad 30, 28911 Legan\'es, Spain} \ead[url]{http://scala.uc3m.es}

\begin{abstract}
A semiclassical model of charge transport in a semiconductor
super\-lattice is solved, using moments in the wavenumber direction
and finite elements in the spatial direction (first order). The
selection of numerical methods guarantees the conservation of current
while allowing for high accuracy results. When a dc voltage bias is held between the ends of the sample, self-sustaining oscillations of the current through the superlattice
are observed in a narrow range of voltages. the calculated solution displayed the
expected accuracy: Spectral convergence in the number of moments used,
and first-order convergence in the number of grid-cells. This result
paves the way for higher-order methods (in the spatial direction) and
the numerical solution of more complex models of charge transport including quantum models based on the Wigner function.
\end{abstract}

\begin{keyword} Semiconductor superlattice \sep kinetic equation of Boltzmann-Poisson type \sep contact
boundary conditions \sep self-sustained current oscillations \sep spectral methods

\PACS 73.63.Hs, 05.60.Gg, 85.35.Be, 02.70.Hm, 02.60.Lj
\end{keyword}
\end{frontmatter}

\nomenclature[um]{$m$}{Subscript denoting the timestep in the
  numerical method}
\nomenclature[ui]{$i$}{Subscript denoting the cell in the numerical
  method. $i=1$ correspond to the first cell, with left boundary at
  $x=0$. $i=N_x$ corresponds to the last cell, with right boundary at $x=L$}

\nomenclature[nl]{$L$}{Dimensionless length of super\-lattice device, $N_p l/x_0$=45}
\nomenclature[ns]{$\varsigma$}{Non-dimensional advection
  coefficient. $\frac{\Delta l}{4 \pi \hbar v_M}=0.582189$}
\nomenclature[na]{$\alpha$}{A pre-factor in the definition of $\fFD$. $\frac{m_*k_BT}{\pi\hbar^2N_D}=0.925115$}
\nomenclature[nt]{$\tilde M$}{The value such that ${\hat f}^\FD_0(\tilde M)=1$, 7.10491}
\nomenclature[nt]{$\tau_e$}{$\sqrt{1+\frac{\nui}{\nue}}$}
\nomenclature[neta]{$\eta$}{The scaled collision frequency, $\oneover{t_0 \nue}=0.476181$}
\nomenclature[ndelta]{$\delta$}{The non-dimensional energy barrier
  height, $\frac{\Delta}{2k_B T}=29.8402$}
 \nomenclature[nbeta]{$\beta$}{The non-dimensional contact conductivity
   $\frac{2\pi\hbar F_M\sigma}{e\Delta N_D}=0.440331$}
\nomenclature[nphi]{$\phi$}{The average electric field, unless stated otherwise, we show
  results for $\phi=1$}
\nomenclature[nNx]{$N_x$}{The number of grid-cells in the numerical simulation}
\nomenclature[nh]{$h$}{The size of each grid-cell. $h=L/N_x$}
\nomenclature[ndt]{$\Delta t$}{The timestep in the numerical method.}
\nomenclature[nNm]{$N_m$}{The number of Fourier modes used in the
  numerics $N_m=2N+1$}
\nomenclature[nN]{$N$}{The index $j$ of the fastest oscillating
  moments in the simulation}
\nomenclature[vx]{$x$}{The physical dimension of the SL,
  non-dimensionalized with units $x_0$}
\nomenclature[vk]{$k$}{The momentum of electrons in the super\-lattice}
\nomenclature[vt]{$t$}{Time, non-dimensionalized with unit $t_0$}

\nomenclature[psigma]{$\sigma$}{The contact conductivity $250 (\meter\usk\ohm)^{-1}$}
\nomenclature[pd]{$\Delta$}{The difference in base energy between the two materials, $72 \milli \electronvolt$.}
\nomenclature[pd]{$d_B$}{The width of the ``barrier'' material, 3.64\nano\meter }
\nomenclature[pd]{$d_W$}{The width of the ``well'' material,
   0.93\nano\meter}
\nomenclature[pnd]{$N_D$}{The density of impurities in the semiconductors, $4.57\times10^{14} \meter^{-2}$}
\nomenclature[per]{$\epsilon_r$}{The relative permittivity of the semi-conductor material, $12.85$}
\nomenclature[pNp]{$Np$}{The number of periods in the super\-lattice, $157$}

\nomenclature[dl]{$l$}{$d_W + d_B$, the period of the super\-lattice, 4.75\nano\meter }
\nomenclature[dm]{$m_*$}{Electron's effective mass $(0.067 d_W + 0.15 d_B)\frac{m_0}{l}=7.64191\times10^{-32} \kilo\gram$}

\nomenclature[sv]{$v_M$}{Scaling unit of electron drift velocity,
  $\frac{\Delta l {\hat f}^\FD_1(\tilde M)}{4\hbar\tau_e}=68.3296 \kilo\meter\per\second$}
\nomenclature[sx0]{$x_0$}{The scaling unit of $x$ in the
  super\-lattice, $\frac{\epsilon_r F_Ml}{eN_D}=15.9439 \nano\meter$}
\nomenclature[st0]{$t_0$}{The scaling unit of time, $\frac{x_0}{v_M}=0.233338 \pico \second$}
\nomenclature[sj0]{$j_0$}{The scaling unit of current density,
  $\frac{e v_M N_D}{l}=1.094761\times 10^9 \ampere\per\meter^2$}
\nomenclature[sF]{$F_M$}{The scaling unit of the electric field
  $\frac{\hbar\sqrt{\nue(\nue+\nui)}}{e l} =
  2.24519\times10^6\joule\per(\coulomb\usk\meter)$}

\nomenclature[ch]{$\hbar$}{$1.05457\times 10^{-34} \joule\usk\second$. Dirac's constant}
\nomenclature[cn]{$\nue$}{$9\times 10^{12} \hertz$, the collision frequency of electrons with each other}
\nomenclature[cn]{$\nui$}{$18\times 10^{12} \hertz$, the collision frequency of electrons with impurities}
\nomenclature[cm0]{$m_0$}{The rest mass of an electron, $9.10938\times10^{-31} \kilo\gram$}
\nomenclature[cel]{$e$}{The charge of an electron, $1.60218\times10^{-19}\coulomb$}
\nomenclature[ce0]{$\epsilon_0$}{The permittivity of vacuum,
  $8.8541\times 10^{-12} \coulomb\per(\meter\usk\volt)$}

\nomenclature[fTheta]{$\Theta(k)$}{The Heaviside function
  $\Theta(k)=1$ for $k\ge0$ and zero otherwise}

\nomenclature[vJ]{$J$}{The mean current density: $J=\oneover{L}\int_0^L j(x,t)\,dx=\oneover{L}\int_0^L \sqrt{\pi}\varphi\hat f_{-1}(x,t)\,dx$}

\section{Introduction}
Bloch oscillations are coherent oscillations of the electron position inside an energy band of a crystal under an applied electric field. Their frequency is proportional to the field times the lattice constant and it should be larger than the inverse scattering time for the oscillations to persist. The necessary electric field is too large for natural crystals and thus Esaki and Tsu suggested in 1970 to construct an artificial crystal with a larger effective lattice constant called a superlattice (SL) \cite{ET1970}. The simplest SL example is formed by epitaxially growing many identical periods comprising a number of layers of two different semiconductor materials \cite{BG2005}. The difference in the energy gaps of the component semiconductors causes the conduction band of the super\-lattice to be a periodic succession of barriers and wells with typical periods of several nanometers. Provided the lateral extension of a SL is much larger than its period, it is a quasi one-dime
 nsional (1D) system. Damped Bloch oscillations of terahertz frequency were first observed in 1992 in such undoped semiconductor SLs whose initial state was prepared optically \cite{FLSMCMPSTS1992}. These SLs had finitely many spatial periods and were subject to an appropriate DC voltage bias. In SLs made out of doped semiconductors, scattering usually destroys the Bloch oscillations but, in theory, they can persist even in the hydrodynamic regime for a SL with long scattering times \cite{BAC2011,ACB2012}. 
 
Except for a narrow parameter range, Bloch oscillations are not stable states in doped SLs  \cite{BAC2011,ACB2012}. However there are other stable self-sustained oscillations (SSCO) of the current that are observed in a DC voltage biased SL. 
These oscillations have frequencies in the gigahertz range and are caused by repeated formation of electric field pulses at the injecting contact of the SL that move forward and disappear at the receiving contact. They have been observed in experiments with \ce{GaAS}/\ce{AlAs} SL (and with other SL based on III-V semiconductors) since 1996 and are the basis of fast oscillator devices \cite{HGSIRPKMISK1996}, which have important applications in industry.

At the most fundamental level, nonlinear transport in SLs has been modeled using quantum kinetic equations based on nonequilibrium Green functions \cite{WACKER02}, Wannier-Stark distribution functions \cite{ROTT99} or Wigner-Poisson equations \cite{BE2005,AB2010}. In the latter case, reduced nonlocal drift-diffusion equations for the electric field and the electron density can be derived using the Chapman-Enskog perturbation method \cite{BE2005,AB2010}. In the semiclassical limit, these equations coincide with those similarly derived for semiclassical Boltzmann-type equations \cite{BEP2003}. Mathematical models at the level of semiclassical kinetic theory go back to the 1970s \cite{KSS1972} but, in the early work, their analysis was based on simplified reduced rate equations (ordinary differential equations) \cite{IS1987, IDS1991} which typically ignore space-charge effects. Electron transport in a single-miniband SL can be described by a semiclassical kinetic equation couple
 d to a Poisson equation approximately describing the electric potential due to the other electrons \cite{BEP2003}. The electron density in the \mbox{$x-k$} space (position and momentum) satisfies a two-dimensional, non-linear, hyperbolic PDE, coupled to a Poisson equation which depends on the average charge density.

Recently, Cebri\'an et al \cite{CBC2009}  numerically solved the kinetic
equation using a direct approach
and showed that self-oscillations are among its solutions and also
studied the relation between these solutions and those of the limiting
drift-diffusion equation.  
Their numerical solution was based on a hybrid particle/fixed-grid
method that used the particles to solve the advective terms and 
used the grid for calculating the solution to the Poisson problem and
for evaluating the effect of the source term.
This has several disadvantages: 
First, it adds a layer of complication as the solution needs to be continually projected back and forth from the particles to
the grid; second, despite the
conservative nature of the equations it has not been shown that the resulting method is conservative. In fact, due to the use of averaging for obtaining point-wise values, there is reason to believe that it is not; lastly, it would be
difficult to extend this solution method to allow for a time-dependent bias
voltage and for solving quantum kinetic equations away from the semiclassical limit.

Here, we solved the charge-transport equation using moments
(Fourier basis) in $k$. 
This approach has several advantages: 
\begin{itemize}
\item The 2-D PDE is transformed into a system of 1-D
  conservation laws (for the coefficients of the moments) which can be
  solved using a standard method (upwind,
  Godunov method); this is indeed how we solve it.
\item The zero moment's (average charge density) equation has no source
  terms, which makes guaranteeing a conservative solution much easier.
\item Since the total density is one of the dependent variables,
 solving the Poisson equation involves only a simple linear equation.
\item The first two moments are the current density and energy
  density, quantities of high physical significance and importance.
\item The resulting method can be generalized to solve more realistic
  models, for example using the Wigner-Poisson quantum kinetic equation\cite{BE2005,AB2010}, a more complete collision model \cite{BAC2011} or a time-dependent voltage bias.
\item Due to the spectral convergence of Fourier expansion, the
  computational cost is lower using this method, for the same accuracy. 
\end{itemize}
Our use of Moment methods is similar to the use thereof in the problem of Radiative Transfer. 
This was first derived formally by Gelbard \cite{Gelbard60,Gelbard61,Gelbard62} and has been used by many since then, for example by Frank et al.\cite{FLS11, FKLY07}. 
The fundamental idea is to write the solution using a family of rapidly converging basis functions and then derive the equations for the coefficients. In the case of the Boltzmann equation for rarefied gas dynamics, spectral methods have been used after constraining the velocities (equivalently, wave numbers) to take values on a bounded domain with periodic boundary conditions and modifying accordingly the collision term \cite{PR2000,FMP2006}. In our case, the wave number takes values on a bounded interval and the distribution function is periodic in it, so that we do not have this additional source of numerical error.

Inevitably, a numerical approximation will require the truncation of the series of basis functions to a finite sum, but the equations describing the evolution of the coefficients will not be ``closed'', that is, it will involve one (or more) of the truncated coefficients. 
An external argument is normally needed in order to ``close'' the resulting equations, and while the zero-closure (assume that the truncated coefficients vanish) is easy to implement and usually good enough (due to the spectral convergence), other closures, such as maximum entropy \cite{JAYNES57} or optimal prediction \cite{FS07}, can be considered as they can lead to significant increase in the accuracy for a given number of kept moments. 

The paper is organized as follows. In Section~\ref{sec:model}, the non-dimensional equations that describe the system are presented; in Section~\ref{sec:moments}, the method of moments is described as it applies to the equations at hand; in Section~\ref{sec:implementation}, the implementation of the numerical method is described. Results and conclusions are presented in sections~\ref{sec:results} and~\ref{sec:conc}. Nomenclature can be found in Section~\ref{sec:nomen}.

\section{Non-Dimensional Model}
\label{sec:model}
As others before \cite{CBC2009, BG2005}, we non-dimensionalize the charge
transport equations using units which, like all other symbols in this
paper, can be found in Section~\ref{sec:nomen}.

The non-dimensional of equations for the electron density $f(k,x,t)$
is:
\begin{equation}
f_t + 2 \pi \varsigma \sin(k) f_x + \frac{\tau_e}{\eta} F(x) f_k =
\oneover{\eta}\brk{\fFD(k,\mu)-f(k)(1+M) + Mf(-k)}.
\nomenclature[vf]{$f$}{The distribution of electrons as a function of
  $x$, $k$, and $t$\nomrefeq}
\label{eq:PDE}
\end{equation}
\nomenclature[nM]{$M$}{The ratio between the collision constants: $M=\frac{\nui}{2\nue}$.}
Here, $f$ and $F$ are coupled via a Poisson equation for the
potential $V$:
\begin{align}
\label{eq:poisson}
V_{xx}&=F_x=n-1, \quad V(0)=0,\quad V(L)=\phi L\\
\nomenclature[vV]{$V$}{The electric potential in the sample\nomrefeq}
\nomenclature[vF]{$F$}{The electric field at a point $x$ in the sample\nomrefeq}
\label{eq:fourier:zero}
 n(x,t)&=\oneover{\sqrt{2\pi}}\hat f_0(x,t)=\oneover{\sqrt{2\pi}}\int_{-\pi}^\pi \cos(j k) f(k,x,t)\, dk\,.
\nomenclature[vn]{$n$}{The total electron density at a point $x$\nomrefeq}
 \intertext{where $\hat f_0$ is the $\nth{0}$ Fourier mode\footnote{We use
     the unitary definition of Fourier modes to keep the problem self-adjoint} of $f$. The others modes are given by:}
\label{eq:fourier:pos}
\hat f_j&=\oneover{\sqrt{\pi}}\int_{-\pi}^\pi \cos(j k) f(k,x,t)\, dk\quad \text{ for } j> 0\,,\text{ and} \\
\label{eq:fourier:neg}
\hat f_{-j}&=\oneover{\sqrt{\pi}}\int_{-\pi}^\pi \sin(j k) f(k,x,t)\, dk \quad\text{ for } {-j}<0\,,
 \nomenclature[ff]{$\hat f_j$}{The $j$th moment of a (non-dimensional)
   electron density  $f$, see
   Eqs.~(\ref{eq:fourier:zero}--\ref{eq:fourier:neg})}
\nomenclature[uj]{$j$}{Subscript denoting the Fourier mode\nomrefeq}
\intertext{while $\fFD$ is the Fermi-Dirac distribution, given by}
\fFD(k,\mu)&=\alpha\log\fprn{1+\exp\fbrk{\mu-\delta(1-\cos(k))}}.
 \nomenclature[ff]{$f^\FD(k;\mu)$}{The Fermi-Dirac distribution\nomrefeq}
\intertext{In the definition of $\fFD$, $\mu=\mu(n)$ is the unique value for which}
n&=\oneover{\sqrt{2\pi}}\hat f^\FD_0(\mu)
\nomenclature[vmu]{$\mu$}{The chemical potential at a given point $x$\nomrefeq}
\end{align}
 \nomenclature[ff]{$\hat f^\FD_j$}{The $j$th moment of the Fermi-Dirac distribution $\fFD(\mu)$}
Where the Fourier modes of $\fFD$ are defined equivalently to those of $f$.
\subsection{Boundary conditions and initial conditions}
To make the problem well-posed, we need to supply initial conditions for $f(x,k,0)$ and
boundary conditions for $f(0,k,t)$ and $f(L,k,t)$. Both require one more definition. 

A steady-state solution at a constant field, $F$, would have vanishing
time- and space-derivatives. We define this distribution as $\fzero$
and use it both in the initial conditions and in the boundary conditions.
Setting the time- and space-derivatives to zero in  \eqref{eq:PDE}, we get a (non-local) ODE for $\fzero(k;F,n)$:
\begin{equation}
\label{eq:f_0}
(1+M)\fzero(k)-M\fzero(-k)+\tau_eF \partial_k\fzero = \fFD(k,\mu(n)),
\nomenclature[ff]{$\fzero(k;F,n)$}{The steady-state distribution for a
  given $F$ and $n$\nomrefeq}
\nomenclature[ff]{$\fzero(k)$}{short-hand for $\fzero(k;F,n)$ with $F$
  and $n$ from the context}
\end{equation}
As shown below, a solution to this equation is straight-forward using Fourier series.
We assume that the initial condition solution is  $\fzero$ with
$n\equiv1$ and $F\equiv\phi$:
\begin{equation}
  \label{eq:f:initial}
  f(x,k,0)=\fzero(k;\phi,1).
\end{equation}
As for the boundary conditions, we expect that the $x=0$ terminal will
be \emph{injecting} electrons, and the $x=L$ terminal will be \emph{collecting} them. 
We therefore follow others in using a ``top-down'' approach and
require that the current at the injecting terminal obeys (dimensional) Ohm's Law $j=\sigma F$, while at
the collecting terminal we simply require that the (dimensional)
electron density is $N_D$. The non-dimensional versions of these BC are:
\begin{equation}
  \label{eq:BC:non-D}
  j(0,t)=2\beta\varsigma F(0),\qquad n(L,t)=1.
\end{equation}
Here, $j$ is the \emph{local current density}:
\begin{align}
  \label{eq:current:density}
 j(x,t)&=\varsigma\int_{-\pi}^{\pi}\sin(k) f(k,x,t)\, dk
 = \sqrt{\pi}\varsigma\hat f_{-1}(x,t)
\end{align}
There are several ways to achieve these requirements. 
Since Eq.~\eqref{eq:PDE} is hyperbolic, we may only set boundary conditions
where the characteristics are going into the domain, that is, for
$x=0$ we should only set conditions for $k>0$, and for $x=L$, we should only set
conditions for $k<0$.
Since we have only one condition for every boundary, the problem is under-determined. 
We deviate slightly from the choice made in \cite{CBC2009} and use a multiple of $\fzero$ as the boundary condition:
\begin{align}
  \label{eq:BC:f:0}
f(0,k>0,t)&=\frac{\fzero(k)}{\int_{0}^{\pi}\sin(k) \fzero(k)\,dk}\prn{2\beta F-\int_{-\pi}^0 \sin(k)f(0,k,t)\,dk}\\
  \label{eq:BC:f:L}
f(L,k<0,t)&=\frac{\fzero(k)}{\int_{-\pi}^0 \fzero(k)\,dk} \prn{2\pi-\int_0^{\pi}f(L,k,t)\,dk}
\end{align}
A quick check shows that with these definitions the BC at $x=0$
and $x=L$ are satisfied.
In the $k-$direction we impose periodic boundary conditions.

In summary, the problem consists of advection PDE~\eqref{eq:PDE} coupled with
the Poisson problem~\eqref{eq:poisson}, initial
condition~\eqref{eq:f:initial}, and boundary conditions~(\ref{eq:BC:f:0}, \ref{eq:BC:f:L}) in the $x-$direction, and periodicity in the $k-$direction.

\section{Method of moments}
\label{sec:moments}
We solve this model using Fourier modes in the $k-$direction and Gudunov method with wave-splitting on a 
regular grid in the $x-$direction. 
This has the advantage of imitating the charge density conservation property 
that the original equations have, and provides us
with the important physical variables (charge density, current
density and energy density) for ``free'', without the need to
calculate them from the solution.

Multiplying Eq.~\eqref{eq:PDE} by $\oneover{\sqrt{\pi}}\sin(jk)$,
$\oneover{\sqrt{\pi}}\cos(jk)$, or $\oneover{\sqrt{2\pi}}$ and
integrating from $-\pi$ to $\pi$ (with respect to $k$) results in the following
system of equations for $\hat f_j$, the Fourier coefficients of $f$:
\begin{alignat*}{4}
 \partial_t\hat f_{j}& +\pi\varsigma \partial_x(\hat f_{-j-1}-\hat
  f_{-j+1})&&+\frac{j\tau_eF(x)}{\eta}\hat f_{-j}&&=\oneover{\eta}\prn{{\hat
      f}^\FD_{j}-\hat f_{j}}&\text{ for } j&\ge2\\
\partial_t\hat f_1 &+\pi\varsigma \partial_x\hat f_{-2}&&+\frac{\tau_eF(x)}{\eta}\hat f_{-1}&&=\oneover{\eta}\prn{{\hat f}^\FD_{j}-\hat f_{j}}&\text{ for } j&=1\\
\partial_t\hat f_0 &+\pi\varsigma \sqrt{2} \partial_x\hat f_{-1}&&&&=\oneover{\eta}\prn{{\hat
  f}^\FD_0 -\hat f_0}=0&\text{
  for } j&=0\\
 \partial_t\hat f_{-1} &+\pi\varsigma \partial_x(\sqrt{2}\hat f_{0}-\hat f_{2})&&+\frac{\tau_eF(x)}{\eta}\hat f_{1}&&=-(1+2M)\frac{\hat f_{-1}}{\eta}&\text{ for } j&=-1\\
 \partial_t\hat f_{j} &+\pi\varsigma \partial_x(\hat f_{-j-1}-\hat
  f_{-j+1})&&-\frac{j\tau_eF(x)}{\eta}\hat f_{-j}&&=-(1+2M)\frac{\hat f_{j}}{\eta}&\text{ for } j&\le-2
\end{alignat*}
We have used integration by parts, and the symmetry with respect to
$k$ of the Fermi-Dirac distribution, $\fFD$.

One advantage of using Fourier moments is that the equation for $\hat f_0$ has no source term. This encodes the fact
that electrons are not created or destroyed, they are only moved from one place to another with a current density $\hat f_{-1}$. 
By solving the system with a conservative numerical method, we are guaranteed that charge is conserved.

These equations are exact so long as we take all of the infinite moments involved. 
But, of course, when implementing this method we can keep only a finite number of moments, therefore in the extremal
equations---for $f_{N}$ and $f_{-N}$---there will be missing terms: $\hat f_{-N-1}$ for $f_N$ and
$\hat f_{N+1}$ for $f_{-N}$. 
This is a standard problem in moment methods, the solution thereof is referred to as ``moment closure''. 
In this paper we use what is known as the ``$P_N$ closure'' which assumes that the missing moments vanish. 
Other closures may be pursued at a later time.

For a finite number of moments, the moment equations form an
advection-reaction PDE (in $t$ and $x$) for the vector of moments ${\bf\hat f}$: 
\nomenclature[vfvec]{${\bf\hat f}$}{The vector of Fourier terms of $f$}
\nomenclature[vfvecn]{${\bf\hat f}^n_i$}{The Fourier terms of $f$ at the $i-$th cell, during the $n$th time-step }

\begin{equation}
\label{eq:hats}
{\bf\hat f}_t+\pi\varsigma A {\bf\hat f}_x= \oneover{\eta}\brc{\tau_e F(x)S_1{\bf\hat f}+S_2{\bf\hat f}+{\bf\hatfFD}(\mu(n(x)))},
\end{equation}
where $A$ is the advection matrix given by~\eqref{eq:A}, ${\bf\hatfFD}$ is the vector of
Fourier coefficients of $\fFD$, and $S_1$ \& $S_2$ are matrices given by~\eqref{eq:S1:S2}.
\nomenclature[vA]{$A$}{The advection matrix of the
  Fourier moments\nomrefeq}

If we take $N$ positive and $N$ negative moments (and a zero moment), we can write the
 of \eqref{eq:hats} as follows:
\begin{equation}
  \label{eq:A}
\prn{
\begin{BMAT}(rc,0pt,17pt){c}{ccccc.cccc}
\hat f_N\\
\vdots \\
\hat f_2\\
\hat f_1\\
\hat f_0\\
\hat f_{-1}\\
\vdots \\
\hat f_{-N+1}\\
\hat f_{-N}\\
\end{BMAT}}_t+
\pi\varsigma\prn{
\begin{BMAT}(rc,17pt,17pt){rrrrr.rrrr}{ccccc.cccc}
&&&& &&& \tikz[baseline=0] \node[matnode](M_N_-N+1){$-1$}; &\tikz[baseline=0] \node[matnode](M_N_-N){$0$};\\
&&&& &&  & &\tikz[baseline=0] \node[matnode](M_N-1_-N){$1$};\\
&&\smash{\scalebox{2}{$0$}}&&&\tikz[baseline=0] \node[matnode](M_2_1){$-1$}; &&&\\
&&&&&  \tikz[baseline=0] \node[matnode](M_1_-1){$0$}; & \tikz \node[matnode](M_1_-2){$1$}; &&\\
&&&&&  \tikz\node[matnode](M2){$\sqrt{2}$}; & & &\\
&&\tikz[baseline=0]\node[matnode](M_-1_2){$-1$};&\tikz\node[matnode](M_-1_1) {$0$};&\tikz\node[matnode](M_-1_0){$\sqrt{2}$};                          &&&&\\
&&&\tikz\node[matnode](M_-2_1){$1$};&&&&&\\
\tikz[baseline=0] \node[matnode](M_-N+1_N){$-1$};&&&                          &&&&\smash{\scalebox{2}{$0$}}&\\
\tikz[baseline=0] \node[matnode](M_-N_N){$0$};&\tikz[baseline=0]
\node[matnode] (M_-N_N-1){$1$};&&&&   &&&\\
\end{BMAT}}
\prn{
\begin{BMAT}(rc,0pt,17pt){c}{ccccc.cccc}
\hat f_N\\
\vdots \\
\hat f_2\\
\hat f_1\\
\hat f_0\\
\hat f_{-1}\\
\vdots \\
\hat f_{-N+1}\\
\hat f_{-N}\\
\end{BMAT}}_x
\begin{tikzpicture}[overlay]
\path[matline] (M_-1_1) -- (M_-N_N);
\path[matline] (M_N_-N) -- (M_1_-1);
\path[matline] (M_2_1) -- (M_N_-N+1);
\path[matline] (M_N-1_-N) -- (M_1_-2);
\path[matline] (M_-N_N-1) -- (M_-2_1);
\path[matline] (M_-N+1_N) -- (M_-1_2);
\end{tikzpicture}
\end{equation}
The dashed lines separate the negative moments from the non-negative ones as a visual aid.
With the same notation for the vector of $\hat f_j$ values, the
matrices $S_1$, $S_2$ are given by
\begin{equation}
  \label{eq:S1:S2}
  S_1=\prn{
\begin{BMAT}(rc,17pt,17pt){rrrr.rrr}{cccc.ccc}
&&&&&& \tikz[baseline=0] \node[matnode](M_N_-N){$N$}; \\
&&&&&&\\
&&&&\tikz\node[matnode](M_1_-1){$1$};&&\\
&&&  \tikz\node[matnode](M_0_0){$0$}; &&&\\
&&  \tikz\node[matnode](M_-1_1){$-1$};& &&&\\
&&&&&&\\
\tikz[baseline=0] \node[matnode](M_-N_N){$-N$};&   &&&&&\\
\end{BMAT}}
\begin{tikzpicture}[overlay]
\path[matline] (M_N_-N) -- (M_1_-1);
\path[matline] (M_-N_N) -- (M_-1_1);
\end{tikzpicture}
\quad
  S_2=-\prn{
\begin{BMAT}(rc,17pt,17pt){rrrr.rr}{cccc.ccc}
\tikz[baseline=0] \node[matnode](M_N_N){$1$}; &&&&&\\
&&&&&\\
&& \tikz\node[matnode](M_1_1){$1$}; &&&\\
&&&1 &&\\
&&&&  \tikz\node[matnode](M_-1_-1){$1+2M$}; &\\
&&&&&\\
&&&&&\tikz[baseline=0] \node[matnode](M_-N_-N){$1+2M$};\\
\end{BMAT}}
\begin{tikzpicture}[overlay]
\path[matline] (M_N_N) -- (M_1_1);
\path[matline] (M_-N_-N) -- (M_-1_-1);
\end{tikzpicture}
\end{equation}
\nomenclature[vS1]{$S_1$}{One of the matrices used to generate the source term\nomrefeq}
\nomenclature[vS2]{$S_2$}{One of the matrices used to generate the
  source term\nomrefeq}

As mentioned earlier, using these moment the Fourier transform of $\fzero$ from Eq.~\eqref{eq:f_0} is easy to find,
as it must solve Eq.~\eqref{eq:hats} with both derivative terms omitted:
\begin{equation}
  \label{eq:f_hat_0:implicit}
  {\bf0} = \tau_e F S_1{\bf\hatfzero}+S_2{\bf{\hatfzero}}+{\bf\hatfFD}(\mu(n)).
\end{equation}
In other words, 
\begin{equation}
  \label{eq:f_hat_0:explicit}
  {\bf\hatfzero}(n;F) = -(\tau_e F S_1+S_2)^{-1} {{\bf\hatfFD}(\mu(n))}.
\nomenclature[ff0hat]{$\hatfzero$}{The Fourier modes of the
  equilibrium distribution $\fzero$\nomrefeq}
\end{equation}
Thus the initial condition is easy to write in Fourier, what about the
boundary conditions?
Since the boundary conditions \eqref{eq:BC:f:0} and \eqref{eq:BC:f:L} depend on Fourier modes of the truncated density function $\Theta(-k) f(0,k)$ and $\Theta(k) f(L,k)$, we reconstruct $f$ at the boundaries from the Fourier coefficients, truncate the resulting function as appropriate, and calculate the Fourier
coefficients of the truncated function. While this isn't very efficient, it only has to be done at the boundaries, and therefore relatively cheap, computationally.

\section{Implementation}
\label{sec:implementation}
The complete problem consisting of linear advection, source terms, and the
coupled Poisson equation are solved using operator splitting, alternating
between an advection step and a source step. 
The electric field is calculated from the solution of the Poisson equation before it is needed in the source term and boundary conditions. 
The source consists of three terms that can be more accurately solved separately than together. 
Therefore, we also use operator-splitting for the source step itself

\subsection{Advection}
To solve the linear advection system, we use wave-splitting following LeVeque's book\cite{LeVeque}.  
This means that the numerical values represent cell averages and at
every time-step, we write the difference between neighboring cells as
a sum of eigenvectors of the advection matrix, and calculate the change to the cell-averages due to upwind advection of these waves\footnote{The time-step is chosen small enough so that waves from neighboring cells cannot interact}. 
This is a first-order approximation that is consistent with the conservation properties of the problem, and is therefore guaranteed to conserve the charge density.

Given a Riemann problem initial condition (constant solution at each cell), each eigenvector of the
matrix $A$ corresponds to a ``wave'' that travels at a
specific speed $\lambda$ (the corresponding eigenvalue). 
These waves can be followed as they travel forward ($\lambda>0$) or backwards  ($\lambda<0$) and
the cell averages can be adjusted accordingly:
\nomenclature[udotplus]{$(\cdot)^+$}{$\max(0,\cdot)$}
\nomenclature[udotminus]{$(\cdot)^-$}{$\min(0,\cdot)$}
\begin{equation}
\label{eq:wavesplitting}
{\bf\hat f}_i^{m+1}={\bf\hat f}_i^m+\frac{\Delta t}{h}\brk{\mathcal{A}^+ \prn{{\bf\hat f}_{i}^m-{\bf\hat f}_{i-1}^m}+
\mathcal {A}^-\prn{{\bf\hat f}_{i+1}^m-{\bf\hat f}_i^m}}
\end{equation}
Where $\mathcal{A}^\pm$ are the right- and left-going advection velocities given by
\begin{equation}
\mathcal{A}^+=R \prn{\Lambda}^+ R^{-1}, \qquad \mathcal{A}^-=R \prn{\Lambda}^- R^{-1}.
\nomenclature[vlambda]{$\Lambda$}{A diagonal matrix of the eigenvalues of $A$
 so that $AR=\Lambda R$}
\nomenclature[vR]{$R$}{The matrix of eigenvectors of $A$ so that $AR=\Lambda R$}
\end{equation}

For stability we keep $\frac{h}{\Delta t}$ smaller than the fastest wave in the system. 
As $N\goto \infty$, the largest eigenvalue of $A$ approaches 1 (this encodes the maximal value of $\sin(k)$) thus, as a CFL condition we use
\begin{equation}
\Delta t=.95 \frac{h}{\pi \varsigma}
\end{equation}

At the boundaries, we need to provide a value of the solution outside the domain. 
Since only the waves that go \emph{into} the domain affect it, we can provide the values
on both the ingoing and outgoing parts of the solution and let the up-winding take care of 
moving the information in the correct direction.

Following Eq.~\eqref{eq:BC:f:0}, for the left boundary, $x=0$, we let a ``ghost'' cell have the value
\begin{equation}
  \label{eq:BC:hat:0}
  {\bf\hat f}_0=\frac{\bf\hatfzero}{{\hat{\mathcal{F}}\brk{\Theta(k)\fzero(k)}}_{-1}}
  \prn{ \frac{2\beta F}{\sqrt{\pi}}-\hat{\mathcal{F}}\brk{\Theta(-k) \hat{\mathcal{F}}^{-1}\brk{{\bf\hat{f}}_1}}_{-1}}
\end{equation}
\nomenclature[fF1]{$\hat{\mathcal{F}}[\cdot]$}{The Fourier operator, resulting in a $2N+1-$vector, see eqs.~(\ref{eq:fourier:pos}--\ref{eq:fourier:neg})} 
Similarly, at the right boundary, $x=L$, we define another ``ghost'' cell
with the solution
\begin{equation}
  \label{eq:BC:hat:L}
  {\bf\hat f}_{N_x+1}=\frac{\bf\hatfzero}
{\hat{\mathcal{F}}\brk{\Theta(-k)\fzero(k)}_0}\prn{\sqrt{2\pi}-\hat{\mathcal{F}}
\brk{\Theta(k){\hat{\mathcal{F}}}^{-1}\brk{{\bf \hat f}_{N_x}}}_0}
\end{equation}
The inverse Fourier operator is defined as:
\begin{equation}
  \label{eq:inv:fourier:def}
g(k)=\hat{\mathcal{F}}^{-1}\brk{\bf \hat g}(k)=\frac{\hat g_1}{\sqrt{2\pi}}+\oneover{\sqrt{\pi}}\sum_{j=1}^N \hat g_{-j}\sin(j k)+ \hat g_j\cos(j k).
\nomenclature[fF2]{$\hat{\mathcal{F}}^{-1}[\cdot]$}{The inverse Fourier operator, resulting in a function of $k$\nomrefeq} 
\end{equation}
\subsection{Sources}
There are three sources terms to contend with and our ability to solve them analytically differs between them. 
The linear ones are each trivial to solve exactly as the eigenvalues of $S_1$ and $S_2$ can be
calculated once in advance. 
Using a spectral decomposition of a matrix $S$:
\begin{equation}
  \label{eq:SD:decomposition}
  RDR^{-1}=S,\quad \text{ where } D \text{ is a diagonal matrix of eigenvalues}
\end{equation}
The solution to 
\begin{equation}
  \label{eq:general:linear:ODE}
    {\bf\hat f}_t=S {\bf\hat f}, \quad \text{ with }  {\bf\hat f}(t)
    \text{ given}
\end{equation}
can be written as 
\begin{equation}
  \label{eq:ODE:solution}
 {\bf\hat f}(t+\Delta t)=R \exp(\Delta t D) R^{-1}  {\bf\hat f}(t).
\end{equation}

The source term that comes from the Fermi-Dirac distribution is
non-linear but while it only depends on $\hat f_0$, the resulting term
is non-zero only for positive $j$ terms.
This implies that it can be solved with a first-order integrator---we use Forward Euler---with no loss of accuracy.

The three sources are put together using
first-order Operator-Splitting: taking first a step with the
non-linear term, then with $S_1$ and finally with $S_2$.
Since the advection is computed to first-order, any effort for
calculating a second-order solution of the source would be mostly lost. 

\subsection{Fermi-Dirac Distribution}
One of the main bottlenecks of the previous papers was in the
inversion of the chemical-potential function $\mu(n)$.
The problem is that this inversion is needed at every
time-step, at every grid-point $x_i$. 
A relatively fast solver can be written using Newton's method, but
even so, the sheer amount of calculating is time-consuming. 
We have elected to use pre-calculation and then evaluation
using a piecewise cubic Hermite interpolant (for $\mu$) and a spline (for the Fourier coefficients of $\fFD$). 
By calculating once the value of $\mu$ (and also ${\bf \hatfFD}$) at sufficient values of $n$ between 0 and 5\footnote{This is a conservative estimate of the possible range of values we will encounter in the simulation.}, 
we can guarantee that the error of the interpolation is smaller than a desired accuracy. 
The evaluation time is a fraction of that when using Newton's method.

The error of the Hermite interpolation of $\mu(n)$ where
$n\in\brk{n_i, n_{i+1}}$ is bounded by
\begin{equation}
  \label{eq:spline:error}
  \text{Error}\le\frac{\norm{\mu^{(4)}(n)}_{\infty,\brk{n_i, n_{i+1}}}}{4!}\prn{n-n_i}^2\prn{n-n_{i+1}}^2
\end{equation}
For small values of $\mu$ the function $\mu(n)$ behaves asymptotically
like $\log(n)$ and this is also where the large fourth derivatives are
found. 
We therefore estimate the fourth derivative of $\mu(n)$ for $n\ll1$ as 
\begin{equation}
  \label{eq:mu:pppprime}
  \mu^{(4)}(n)\approx \frac{2}{n^3} \text{ for } n\ll1
\end{equation}
We use this to estimate the accuracy of our approximation of $\mu$ and when
it is not accurate enough, we perform a few Newton iterations until the
desired accuracy is achieved (starting, of course, with the
interpolation result). Once the correct values are found, the Hermit interplant is 
modified by dividing the offending segment into two parts, thus greatly increasing the accuracy of the interpolation on it. This process was continued until the error was less than $10^{-12}$ for the function $\mu(n)$.

To evaluate $\hatfFD(\mu)$ we use a spline interpolant on 10000 points between 0 and 10.

\subsection{Poisson equation}
The solution to the Poisson equation is required in order to find the electric field, F, needed in equation \eqref{eq:PDE} and in boundary conditions (\ref{eq:BC:f:0}, \ref{eq:BC:f:L}). 
When using the moment methods, the field will be required in the center of each field, to coincide with the other variables. As the field is a first derivative of the electric potential (the solution of the Poisson equation) it would be optimal if the potential were given on the boundaries of the cells rather than the centers. 
This would also integrate the boundary conditions of the Poisson problem most easily, since they too are given on the boundaries. The only problem with this approach is that we have $N_x$ equations (one for each cell) and only $N_x-1$ values with which to satisfy them (from the values of the potential at the internal edges). 
To find a good candidate for this over-constrained problem, we use a finite element approach.

Our problem therefore is to find the electric potential, $V(x)$, and electric field, $F$, by solving \eqref{eq:poisson}:
\begin{equation}
  \label{eq:Poisson:2}
  V_{xx}=n-1,\quad V(0)=0,\quad  V(L)=\phi L
\end{equation}
for a given $n(x)$ and $\phi$. 
Since our first-order advection solver assumes a piecewise constant solution space, $n(x)$ is assumed constant within each x-cell. 
We can therefore write it as 
\begin{equation}
  \label{eq:n_constant}
  n(x)=\sum_{i=1}^{N_x}n_i \mathbbm{1}_i,\qquad \mathbbm{1}_i(x)=\begin{cases} 
1 & \text{ if } \abs{x-(j-\frac12)h}\le hx \\
0 & \text{ otherwise,}
\end{cases}
\end{equation}
where $h$ is the resolution of the x-grid: $h=L/N_x$, and $n_i$ is the constant value of the charge density in the $i-$th cell.

We separate the solution into two parts, one that satisfies the
boundary conditions and the homogeneous equation $V_{xx}=0$ and another that
satisfies the inhomogeneous equation, but homogeneous BC. 
The first is, of course, $x\phi/L$. 
It is the second part for which finite elements are used.
The elements we use for the potential are triangular $\psi_j$ with \mbox{$j=1...N_x-1$}:
\begin{equation}
\label{eq:v_j}
\psi_j(x)=\begin{cases} 
0 & \text{ if } \abs{x-jh}\ge h x \\
1-\frac{\abs{x-jh}}{h} & \text{ otherwise.}
\end{cases}
\end{equation}
To find the field $F$, we use the following program: 
\begin{enumerate}[1.]
\item Write the potential $V$ as linear combination of the $v_j$: 
\begin{equation}
V(x)=\sum_{i=1}^{N_x-1}
v_i \psi_i(x).
\label{eq:V_linear}
\end{equation}
This gives us $N_x-1$ coefficients, $v_i$, such that for any choice the resulting potential satisfies the boundary conditions $V(0)=V(L)=0$. 
\item  Write the Poisson equation \eqref{eq:Poisson:2} using the density as in \eqref{eq:n_constant} and the potential as in \eqref{eq:V_linear}. Multiply by $\phi_k$ and integrate by parts:
  \begin{equation}
    \label{eq:poisson:parts}
    \int_0^L \sum_{i=1}^{N_x-1}
-v_i \psi_{i,x}(x) \psi_{k,x}(x)\,dx = \int_0^L \prn{\sum_{i=1}^{N_x}(n_i-1) \mathbbm{1}_i(x) }\psi_{k,x}(x)\,dx.
  \end{equation}
\item Switch the order of the integral and the sum on both sides of the inequality. This results in a simple linear equation whose solution is the finite elements solution to the Poisson problem:
\begin{equation}
-S \vec v = M( \vec n-\vec{\mathbbm{1}} ),
\end{equation}
The vectors $\vec v$ and $\vec n$ are the vectors of coefficients $v_i$ and $n_i$, and 
the two matrices, $S$ and $M$ are given by: 
\begin{equation}
 S=\frac{1}{h}
\begin{tikzpicture}[>=latex]
\matrix (S)[matrix of math nodes, left delimiter =(, right delimiter %
=),column sep =0pt] {%
 2 & -1\\
-1 \\
&&&-1\\
&& -1&2\\
};
\draw[matline] (S-1-1) -- (S-4-4);
\draw[matline] (S-2-1) -- (S-4-3);
\draw[matline] (S-1-2) -- (S-3-4);
\end{tikzpicture}
,\quad M=\frac{h}{2}
\begin{tikzpicture}[>=latex]
\matrix (M)[matrix of math nodes, left delimiter = (, right delimiter %
= )] {%
 1 & 1\\
\phantom{1}& \\
&&\phantom{1}&\phantom{1}\\
&& &1&1\\
};
\draw[matline] (M-1-1) -- (M-4-4);
\draw[matline] (M-1-2) -- (M-4-5);
\end{tikzpicture}
\end{equation}
Here $S$ is a square matrix of size $(N_x-1)$, while $M$ has $N_x-1$ rows, and $N_x$ columns (implemented as sparse matrices). These matrices arise from the integrals in \eqref{eq:poisson:parts}
The complete solution (including the boundary conditions) can thus be written as 
\begin{equation}
  \label{eq:poisson:sol}
 \vec v =\vec x\phi/L- S^{-1}M(\vec n-\vec{\mathbbm{1}}).
\end{equation}
where $x_i=i h$ and $\vec{\mathbbm{1}}$ is a vector of ones.
\item To calculate the field, $F$, we numerically differentiate the potential, $V$, thus obtaining the field in the middle of each cell.  This is what we need for the source term. At the boundaries, we extrapolate from the two closest values of $F$.
\end{enumerate}
The result is equivalent to solving using a  simple divided differences approach, with the effective density $n$ at each edge equal to the average of the two densities in the neighboring cells. 
\section{Results}
\label{sec:results}
\begin{figure}
\centerline{
  \resizebox{\textwidth}{!}{\includegraphics{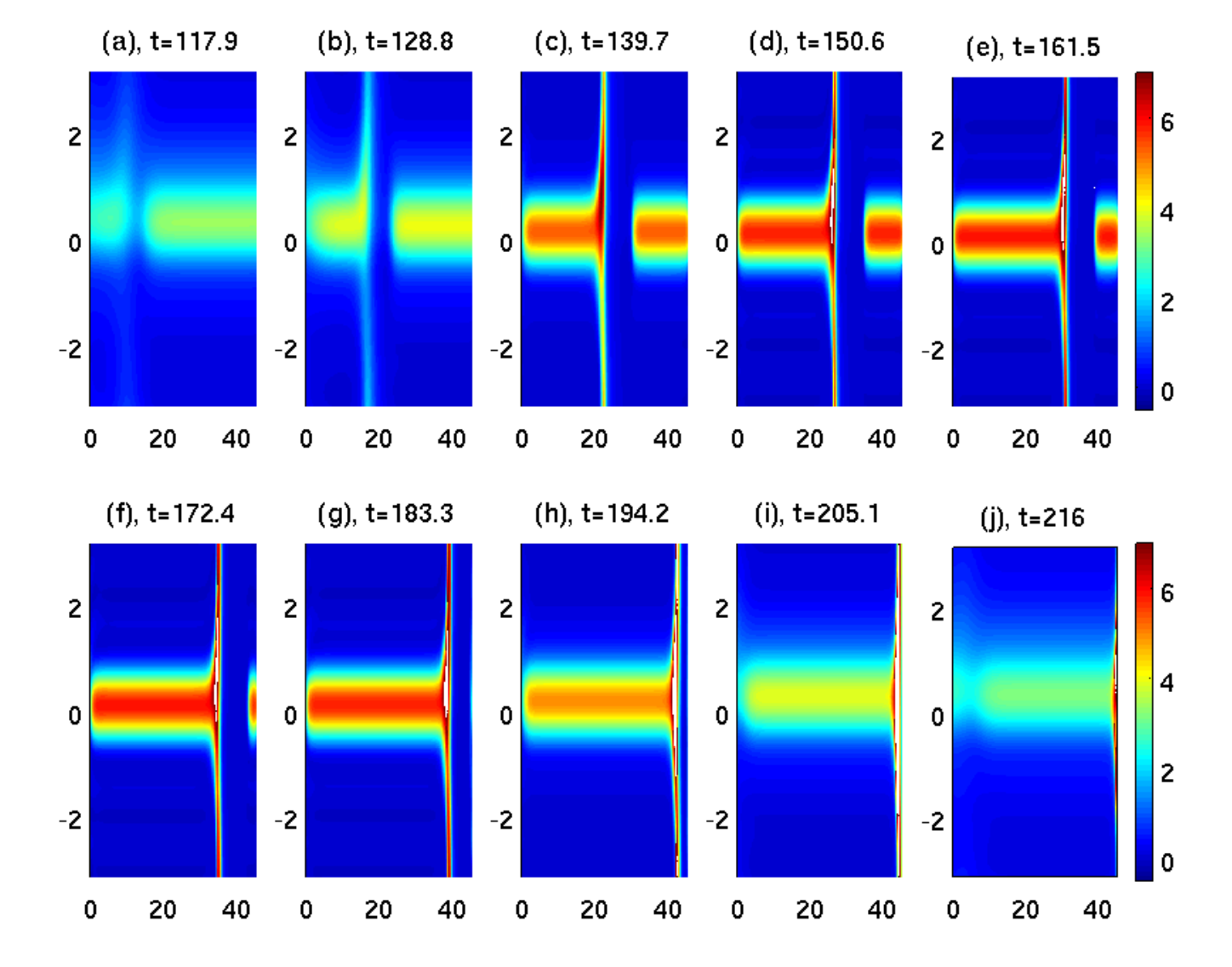}}}
  \caption{The electron density $f(x,k,t)$ during one period of the solution.}
  \label{fig:density:f}
\end{figure}

The method described in the previous section exhibits self-sustained
oscillations very similar to the ones found in
the Cebri\'an paper \cite{CBC2009}, for average bias field $\phi \ge1$
(see Fig.~\ref{fig:density:f}). Specifically, we see that the resulting charge distributions are similar, and that the resulting current densities shows self-sustained oscillations that have similar temporal frequency and similar range. 

\begin{figure}
\centerline{\resizebox{.9\textwidth}{!}{\includegraphics{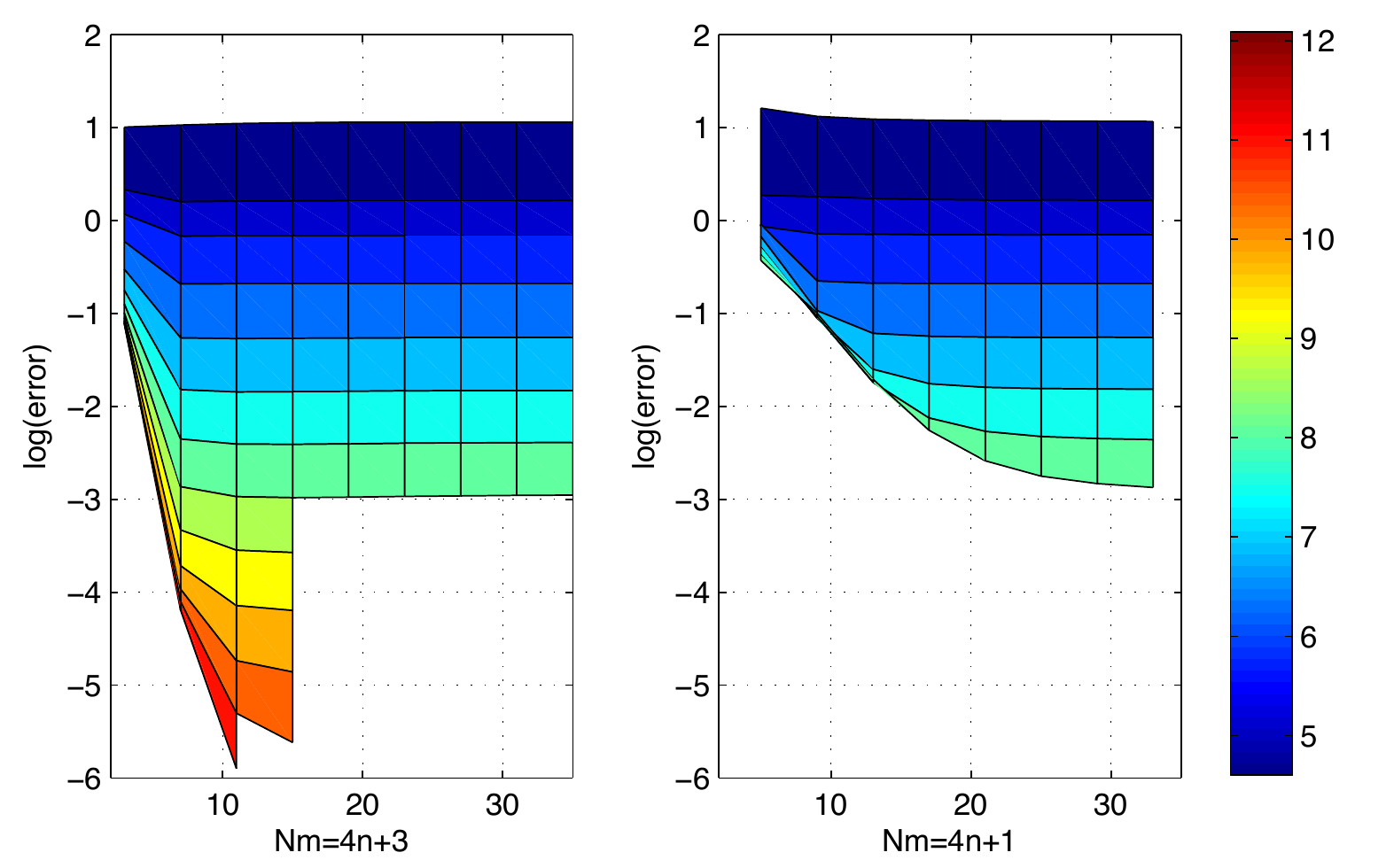}}}
  \caption{The accuracy of the method as a function of $N_x$ and $N_m$. On the left $N_m=3$ (mod 4) and on the right $N_m=1$ (mod 4). The color indicates $log(N_x)$ as per the color-bar on the right ($N_x$ was taken to be the integer part of $10^{k/4}$ with $k=8\ldots20$. }
  \label{fig:convergence}
\end{figure}

We calculate the error by comparing $j$, the current density \eqref{eq:current:density}, obtained at different choices of $N_x$ and $N_m$ with one obtained at $N_m=15$ and $N_x=177827$. 
The $l^2$ distance between the current densities is the reported error. 
Both subfigures in Fig.~\ref{fig:convergence} show that the method displays the expected convergence:
First-order convergence in the number of x-cells and spectral-convergence in the number of moments. 
However, due to the spectral convergence as the error due to the truncation of moments is quickly over-shadowed by the error proportional to $h$ (as evidenced by the horizontal plateau.  
This implies that a second-order implementation of the advection and operator splitting would likely result in higher-order solution in the number of x-cells, and therefore with much more accurate results for the same computation effort.

The same figure also evidences an unexpected dependence on $N_m$.
It is not monotone: the accuracy of $N_m$ that are 1 (mod 4) is much lower than that of those that are 3 (mod 4), though both subsequences seem to have the same limiting plateau as $N_m\goto\infty$. 
Other problems that have been solved using moment methods also exhibit such non-monotone convergence, see for example Davison's book \cite{DAVISON57}. 
In those cases the cause of the poor accuracy is the existence of a zero eigenvalue in the advection matrix. 
To avoid the lower accuracy approximations other authors used even number of moments in their calculations. 
In this paper, we used odd number of moments due to the inherent symmetry of the problem and the physical interpretation of the zeroth moment (charge density). 
Therefore, all our simulations have a zero eigenvalue. 
The difference between 1 (mod 4) and 3 (mod 4) is not in the existence of the zero eigenvalue; 
an examination of the eigenvectors that correspond to the vanishing eigenvalue shows that in the 1 (mod 4) the appropriate eigenvector has a non-vanishing $\hat f_0$ term, while in the 3 (mod 4) the $\hat f_0$ term vanishes. 
We \emph{conjecture} that the error accumulation in the $\hat f_0$ term due to the zero eigenvector is responsible for the higher error in the 1 (mod 4) cases.
The $\hat f_0$ term is more important than the rest since many parts of the problem depend on it. 

We compared the results when using the original boundary conditions set out in \cite{CBC2009} and found very little resulting difference. This affirms the claim that as long as the physical constraints are satisfied (Ohm's law at the emitting terminal and charge neutrality at the collecting terminal) the specific details of the BC are not very important.

\begin{figure}
\begin{center}
\resizebox{.8\textwidth}{!}{
\includegraphics{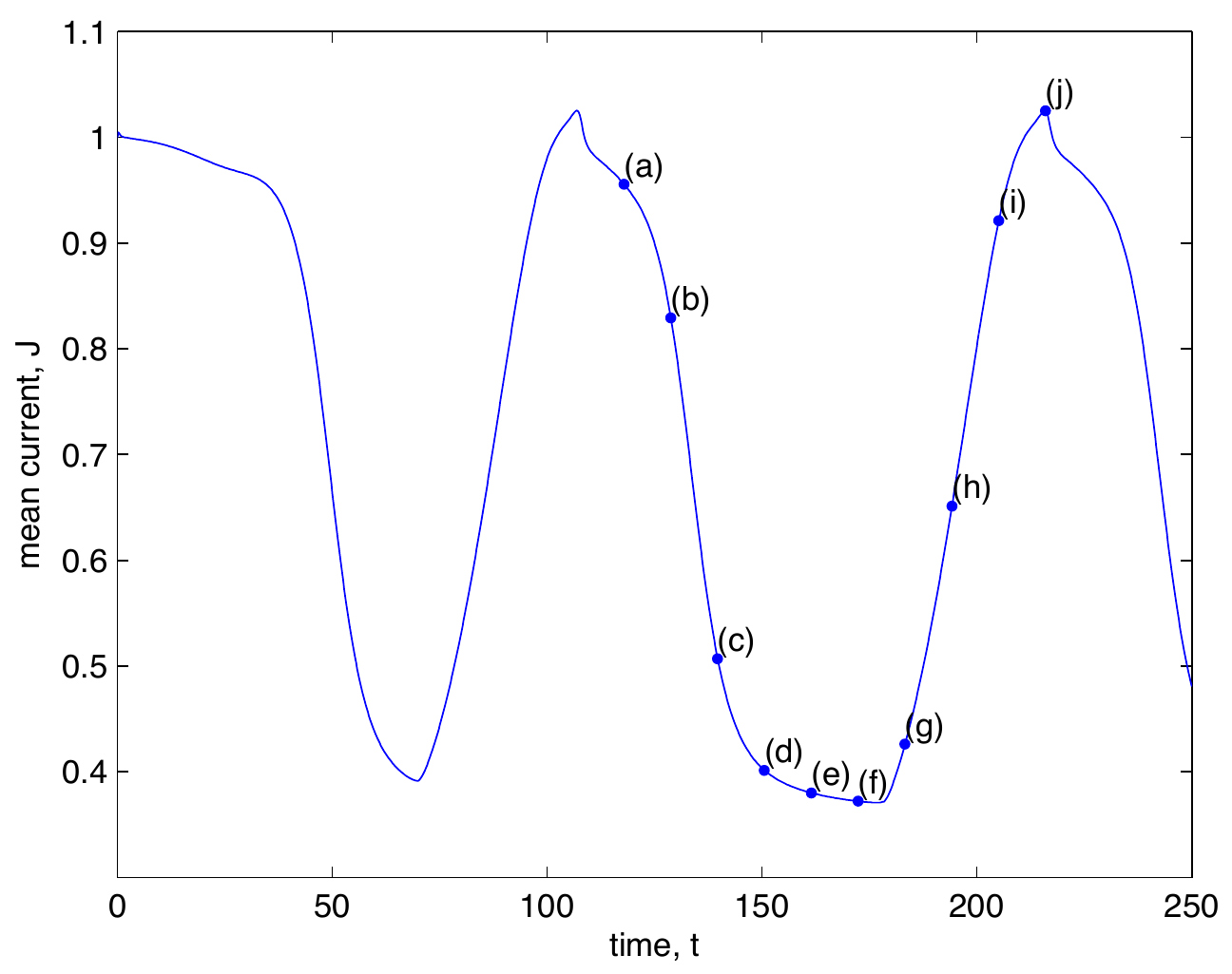}
}
\end{center}
\caption{The mean current  $J(t)$ during the transition and two self-sustaining cycles. The period of oscillation is $~100 t_0$. The markers indicate locations of ``snapshots'' shown in the following figures.}
  \label{fig:current}
\end{figure}
\begin{figure}
\centerline{ \resizebox{.8\textwidth}{!}{\includegraphics{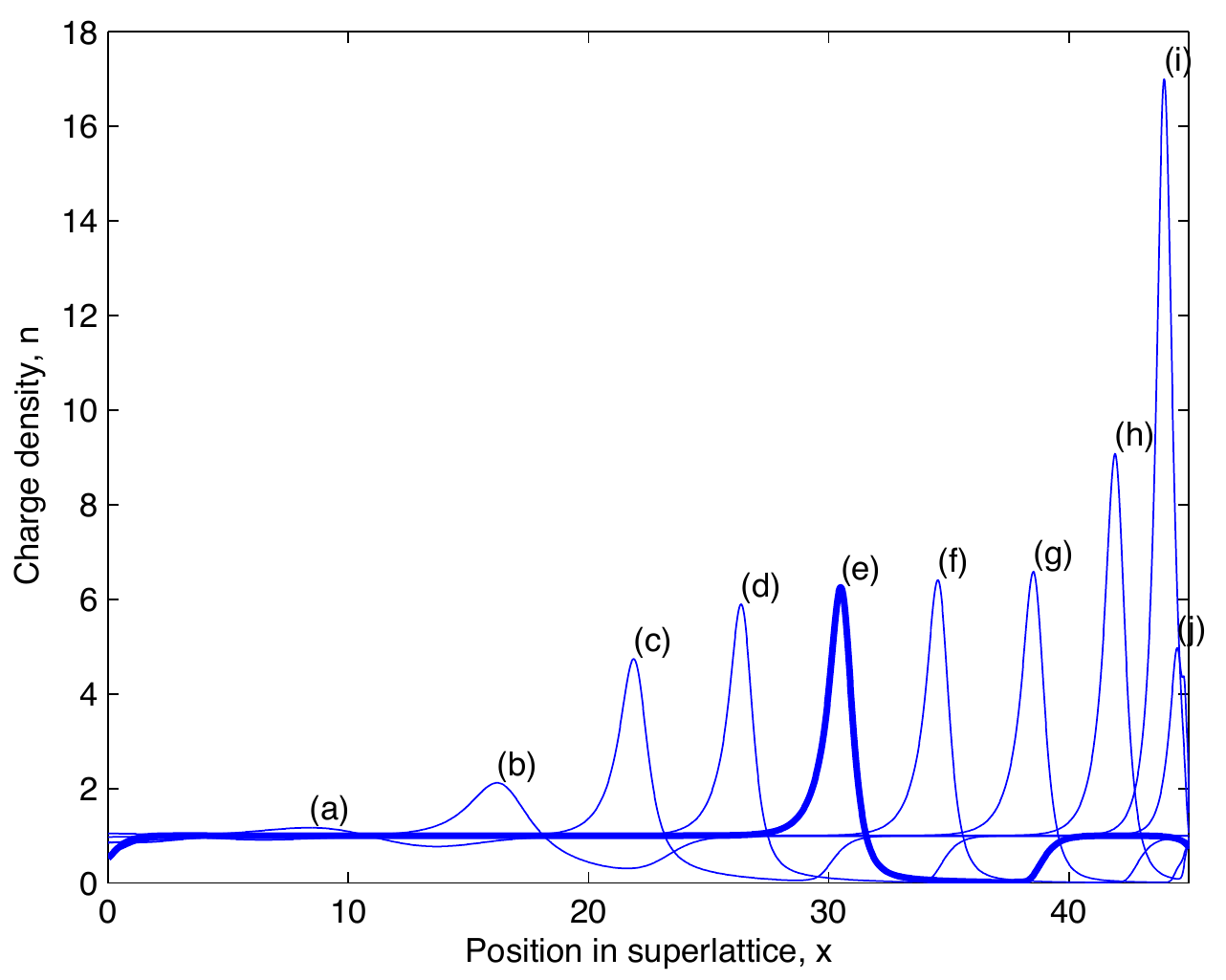}}}
  \caption{The charge density $n(x,t)$ during one period of the solution.}
  \label{fig:density:n}
\end{figure}
\begin{figure}
\centerline{\resizebox{.8\textwidth}{!}{\includegraphics{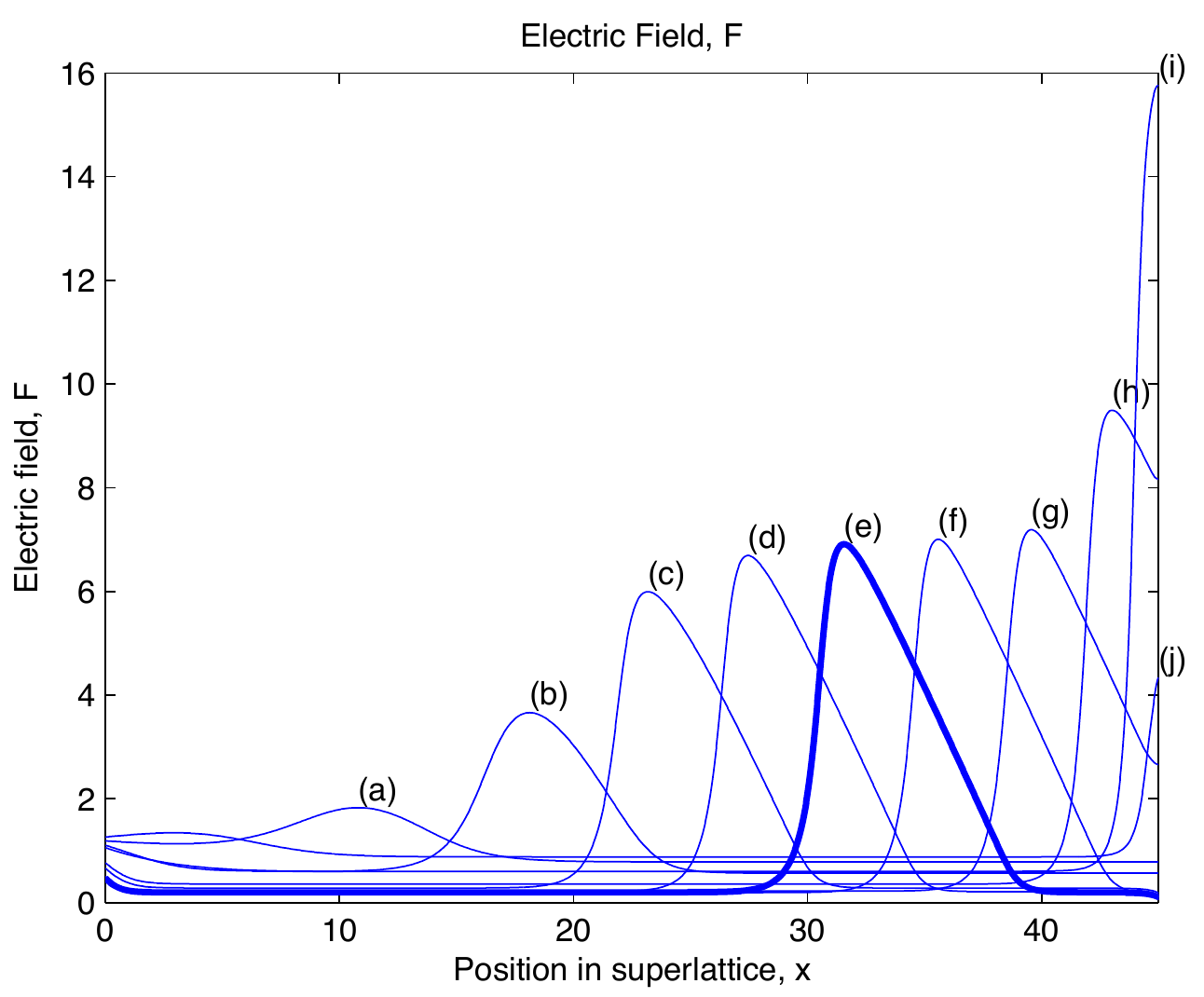}}}
  \caption{The electric field $F(x,t)$ during one period of the solution.}
  \label{fig:field}
\end{figure}
\begin{figure}
\centerline{\resizebox{.8\textwidth}{!}{\includegraphics{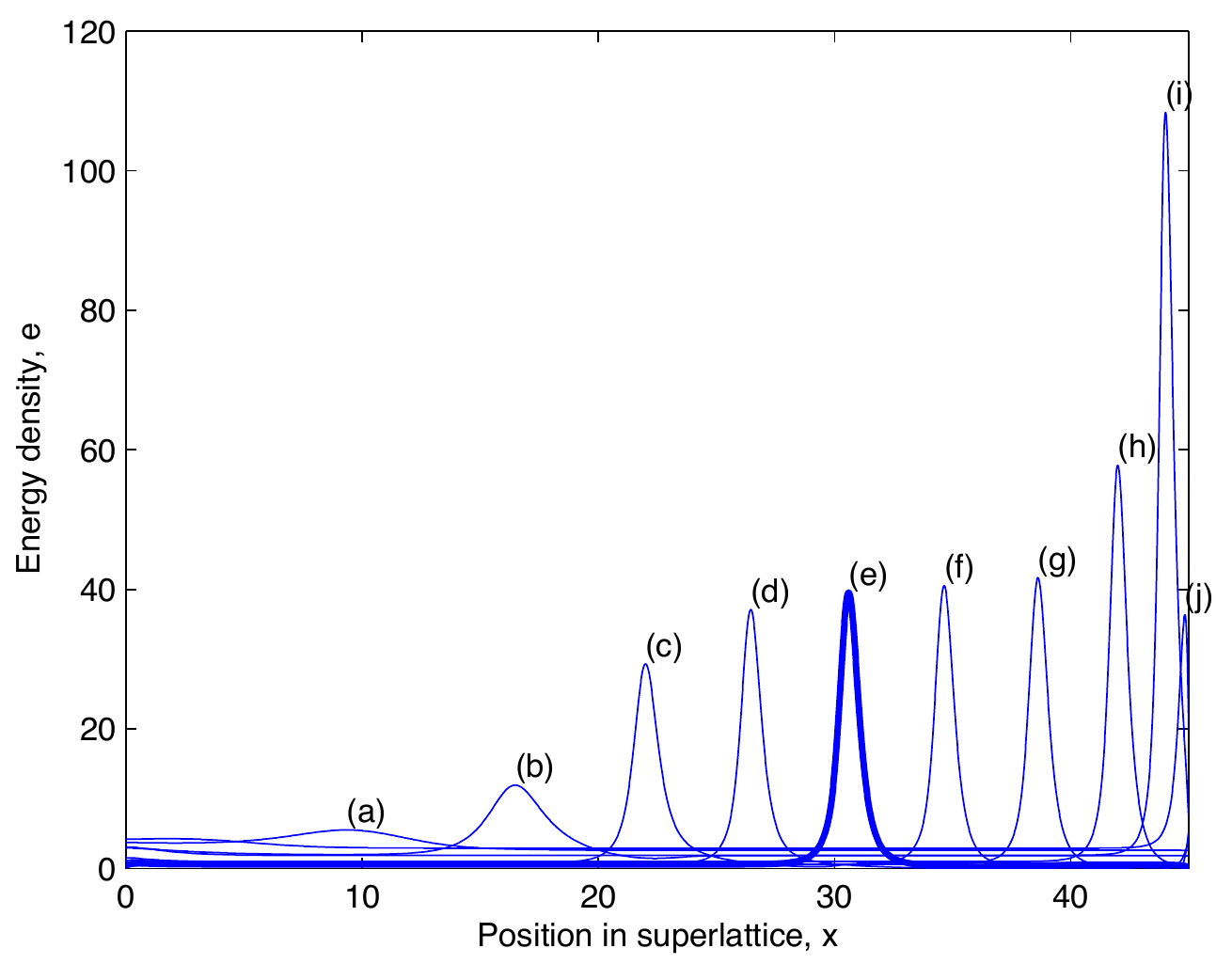}}}
  \caption{The energy density during one period of the solution.}
  \label{fig:energy}
\end{figure}

\begin{figure}
\resizebox{.95   \textwidth}{!}{\includegraphics{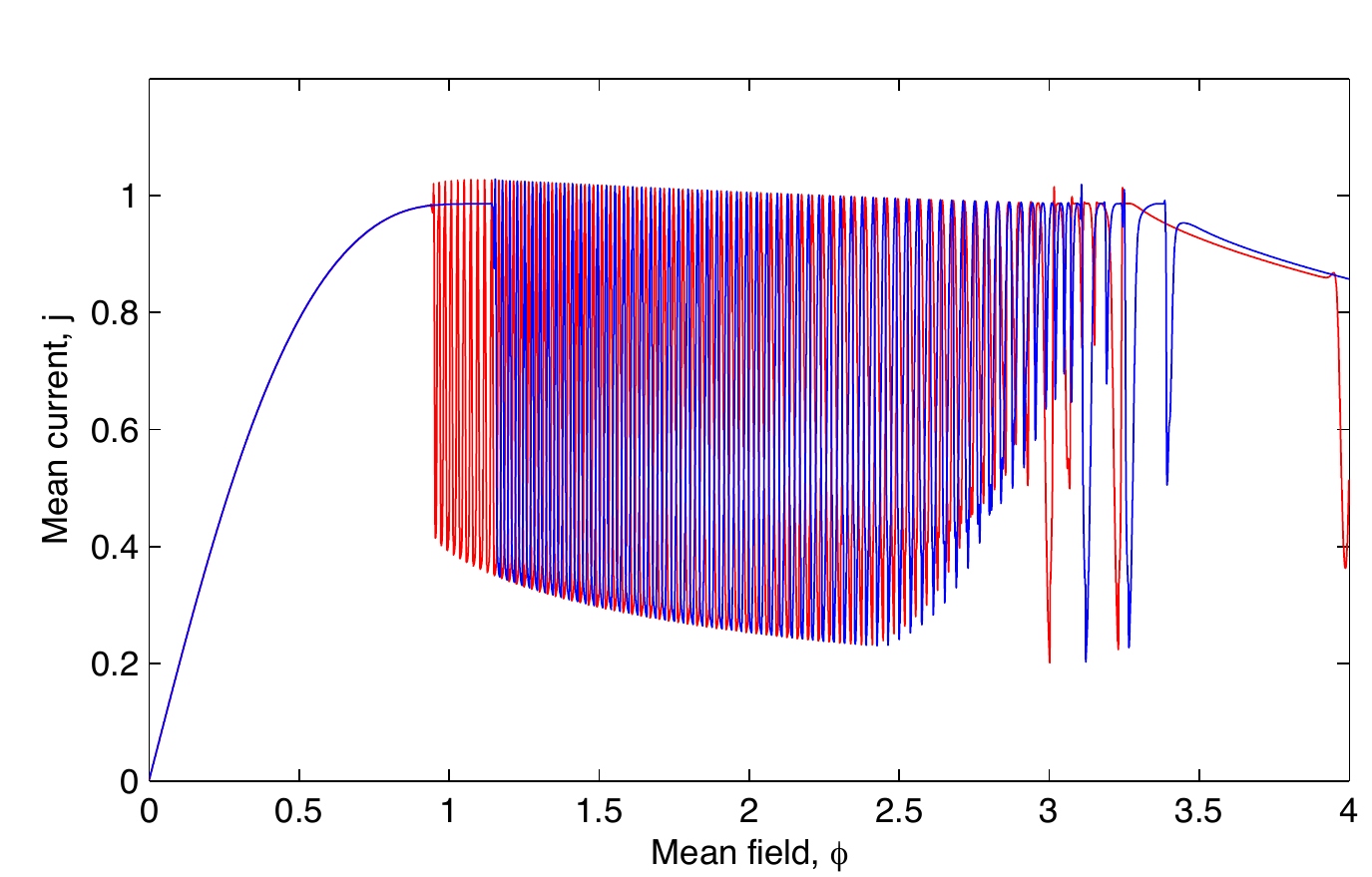}}
  \caption{The mean current during a voltage sweep. In blue the voltage was increasing, and in red---decreasing.}
  \label{fig:hysteresis}
\end{figure}

To study the stability of the system and its response with other values of the voltage, $\phi$, we did two slow (non-dimensional time $t=20000$) ``sweep'' with $\phi$ varying from 0 to 4 and back. 
The resulting mean current is shown in Fig.~\ref{fig:hysteresis}. The results show that in the region $.92<\phi<1.12$ the system can sustain \emph{either} a constant solution or a self oscillating one. This behavior is consistent with a scenario in which the stable self-oscillations appear as a subcritical bifurcation from the stationary state at a critical bias in the previous region. For the related drift-diffusion model of the Gunn effect, such a scenario is realized when the nondimensional length is large enough \cite{BH1995,KHB1996}. This could have interesting applications, as one may be able to encourage the system to pick one behavior over the other by external stimulus.

For $\phi<.92$ the system does not sustain self-oscillations, while for $\phi>1.12$ it not only sustains them, but the constant solution becomes unstable (sub-critically). For even larger values of $\phi$ ($\phi\sim 3)$ the behavior seems erratic, and the high derivatives encountered at the collecting terminal put in question the validity of the results. We intend to repeat these test with a second-order solver to verify. It seems that at these high values of $\phi$ once again only the constant solution is possible, but the accuracy of the solution deteriorates at such large values of $\phi$ due to the resulting high $x-$derivatives of the solution. The results do not provide a clear determination of whether the transition is sub- or super-critical. 

\section{Outlook and Conclusions}
\label{sec:conc}
This paper is a proof of concept---showing a solution to a model of charge transport in a super\-lattice using moments. We used the simplest approaches whenever possible, for example we implemented the integrator with only first order accuracy in the $x-$direction. This allows one to seek improved accuracy in a future study, while knowing that the results, self-sustained current oscillations, do not \emph{depend} on the higher-order method.

Following are several  ways one could improve the accuracy of solution and provide a more complete solution to the charge transport problem:
\begin{itemize}
\item Use a second-order solver for the advection, Poisson problem, and operator splitting.
\item Use a different moment-closure model. 
For example one could use the high moments of $\fzero$ or $\fFD$ to close the moment equations.
\item It may be possible to calculate the maximum entropy moment closure. This implies finding the most likely distribution given the lower moments, and using the moments of \emph{that} distribution to close the equations.
\end{itemize}

We have shown that moment methods can be used to solve the problem of change transport in a super\-lattice. The main difficulties of the original problem (namely, the non-local character of the collision kernel and the integro-differential character of the Poisson problem) are neatly diffused by using a Fourier basis in the $k-$direction. It provide the charge density (required for the Poisson problem) as a dependent variable, enforces the periodic boundary condition in $k$ naturally, and cleanly transforms the non-local collision term into a simple matrix multiplication. The resulting method can be adapted to accommodate other terms that may appear in a less primitive model, and could also be improved by smarter integration methods and moment closure.

\section{Acknowledgments}
This work has been supported by the Spanish Ministerio de Econom\'\i a y Competitividad grant FIS2011-28838-C02-01.
YF was funded as a Juan de la Cierva Researcher at the Universidad Carlos III de Madrid. YF would also like to thank the mathematics department at MIT in which part of this work was done. LLB thanks M.P. Brenner and the School of Engineering and Applied Sciences for hospitality during a stay at Harvard University in which part of this work was done, and a Fundaci\'on Caja Madrid mobility grant for support. The authors would like to thank B.~Seibold for fruitful discussions on moment methods. 

\section{Nomenclature}
\label{sec:nomen}
\printnomenclature

\bibliographystyle{amsalpha}
\bibliography{ChargeTransportWithMoments}

\end{document}